# Touching is believing: interrogating halide perovskite solar cells at the nanoscale via scanning probe microscopy

Jiangyu Li[1,2,*], Boyuan Huang[2], Ehsan Nasr Esfahani[2], Linlin Wei[3], Jianjun Yao[3], Jinjin Zhao[4], and Wei Chen[5]

[1] Shenzhen Key Laboratory of Nanobiomechanics, Shenzhen Institutes of Advanced Technology, Chinese Academy of Sciences, Shenzhen 518055, Guangdong, China

[2] Department of Mechanical Engineering, University of Washington, Seattle, WA 98195, USA

[3] Anasys Instruments, Santa Barbara, CA, USA

[4] School of Materials Science and Engineering, Shijiazhuang Tiedao University, Shijiazhuang, 050043, China

[5] Wuhan National Laboratory for Optoelectronics, Huazhong University of Science and Technology, Wuhan, China

## Abstract

Halide perovskite solar cells based on $CH_3NH_3PbI_3$ and related materials have emerged as the most exciting development in the next generation photovoltaic technologies, yet the microscopic phenomena involving photo-carriers, ionic defects, spontaneous polarization, and molecular vibration and rotation interacting with numerous grains, grain boundaries, and interfaces are still inadequately understood. In fact, there is still need for an effective method to interrogate the local photovoltaic properties of halide perovskite solar cells that can be directly traced to their microstructures on one hand and linked to their device performance on the other hand. In this perspective, we propose that scanning probe microscopy techniques have great potential to realize such promises at the nanoscale, and highlight some of the recent progresses and challenges along this line of investigation toward local probing of photocurrent, work function, ionic activities, polarization switching, and chemical degradation. We also emphasize the importance of multi-modality imaging, in-operando scanning, big data analysis, and multidisciplinary collaboration for further studies toward fully understanding of these complex systems.

[*] To whom the correspondence should be addressed to; Email: jjli@uw.edu

**Introduction**

Since its first report in 2009 [1], halide perovskite solar cells based on $CH_3NH_3PbI_3$ and related materials have emerged as the most exciting development in the next generation photovoltaic technologies [2,3], with its conversion efficiency rising from 3.81% measured in laboratory to 22.1% that is rigorously certified in just 7 years. Such rapid advance in photovoltaic performance is unprecedented in other types of solar cells, fueling intense competition for its efficiency record worldwide. The pace for the new performance record, however, has slowed down considerably since 2016, as 22.1% is already commercially viable and competitive, and much of the recent research focus has been shifted toward more fundamental understanding of photovoltaic characteristics of halide perovskites on one side [4,5], and achieving longer term stability of halide perovskite solar cell structures and performances that is critical for their commercialization on the other side [6–9]. It is within this context that we call for more research efforts into functional probing and imaging of halide perovskite solar cells at the nanoscale.

The photovoltaic performances of a solar cell and its constituents have been conventionally investigated macroscopically at the device level, through characteristics such as transmission spectrum, photoluminescence, current-voltage (IV) curve, quantum efficiency, and impedance, which are global quantities measured with effects of local heterogeneity averaged out. Microscopically, the composition, phase, and structure of photovoltaic materials can be mapped via a number of imaging and analytic techniques with atomic resolution, shedding light into the microstructural mechanisms that are responsible for the macroscopic performance. Nevertheless, there still lacks a critical link between the local structure and global performance, as there still lacks an effective method to interrogate the local photovoltaic properties of solar cells that can be directly traced to their microstructures. For a system as complex as halide perovskite solar cells with interacting photo-carriers, ionic defects, spontaneous polarization, and molecular vibration and rotation on top of numerous grains, grain boundaries, and interfaces, such correlation is essential for understanding their macroscopic photovoltaic characteristics, device efficiency, and degradation. Here we propose that scanning probe microscopy techniques have great potential to realize such promise at the nanoscale, and highlight some of recent progresses and challenges along this line of investigation.

**Photovoltaic Scanning Probe Microscopies**

Scanning probe microscopy (SPM) techniques utilize a cantilever with sharp tip to interrogate the mechanical, electrical, thermal, and chemical responses of the material surface under either local stimuli imposed by the scanning probe or global excitation applied externally [10,11], as schematically shown in Fig. 1, wherein a focused laser can also be used to excite the sample sandwiched between a $TiO_2$ electron transport layer and a 2,2',7,7'-Tetrakis(N,N -di-p -methoxyphenylamino)-9,9'-spirobifluorene (Spiro-OMeTAD) hole transport layer. More information on such device architecture can be found in a recent review [12]. The cantilever functions in a similar way as the human finger: by touching and scanning the sample of interests, not only the surface roughness can be sensed, but also a range of other mechanical and functional characteristics, such as hardness, adhesion, temperature, and charging state, among others, which are usually out of reach for either optic or electron microscopies. With a sharp tip radius unmatched by human finger and well below the optic diffraction limit, spatial resolution as small as a few nanometers can be achieved, making it possible to correlate the local responses directly with microstructural details. Here, we highlight a few SPM studies that have shed light into photovoltaic characteristics of halide perovskite solar cells, and discuss their promises and challenges.

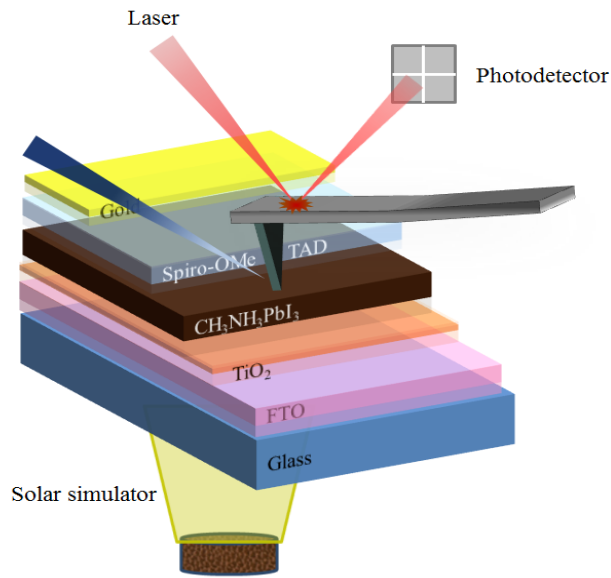

**Fig. 1** Schematics of scanning probe microscopies for perovskite solar cells, wherein $TiO_2$ film is used as the electron transport layer, and 2,2',7,7'-Tetrakis(N,N -di-p -methoxyphenylamino)-9,9'-spirobifluorene (Spiro-OMeTAD) film is used as the hole transport layer; details of such solar cell architecture can be found in a recent review [12].

## *Photocurrent and local IV characteristics*

Conductive atomic force microscopy (cAFM) is capable of mapping local current distribution with sensitivity as high as picoampere (pA), and when such experiments are carried out under light illumination, photoconductive atomic force microscopy (pcAFM) is realized [13]. It has been used to correlate the local photocurrent characteristics with microstructural features such as grains and grain boundaries, for example showing enhanced charge extraction at grain boundaries [14]. In order to obtain IV curves locally at the nanometer scale, point-wise spectroscopy studies can be carried out by sweeping DC voltage biased via bottom electrode while measuring corresponding current, revealing higher hysteresis at grain boundaries as well [15]. Such spectroscopy measurements are rather time consuming, especially for a high density grid of points to achieve desired spatial resolution. Alternatively, photocurrent distribution can be probed under a series of DC biases [16], resulting in a series of current mappings that can be reconstructed to yield mappings of short-circuit photocurrent, open-circuit photovoltage, and fill factor, from which the local photovoltaic efficiency may be mapped with high spatial resolution as well. An example of pcAFM mapping of $CH_3NH_3PbI_3$ on transparent poly(3,4-ethylenedioxythiophene):polystyrenesulfonate (PEDOTS:PSS) substrate spin-coated on fluorine-doped tin oxide (FTO) glass [17] is shown in Fig. 2, and regions of higher and lower photocurrent magnitudes are observed (Fig. 2(a)), reflecting structural heterogeneity of the material. Furthermore, with increased DC biases, the photocurrent gradually drops to zero, as expected, and by averaging over higher and lower current regions or the whole area scanned, representative IV curves can be obtained (Fig. 2(b)), from which the open-circuit photovoltage and short-circuit photocurrent density can be estimated in good agreement with macroscopic measurement (Fig. 2(c)). If the scan covers sufficiently large area that captures the representative variation of microstructural heterogeneity in the sample, then the effect of microstructure on the macroscopic performance can be quantitatively analyzed.

While powerful in correlating photocurrent characteristics with the microstructure, cAFM and pcAFM is only pseudo-local in nature, as the current flow is inherently global, across a path spanning from the conductive probe through the sample to the current collector at the bottom. To truly understand the local characteristics of photocarriers, alternative techniques such as Kelvin probe force microscopy (KPFM) and electrostatic force microscopy (EFM) would be desirable [18].

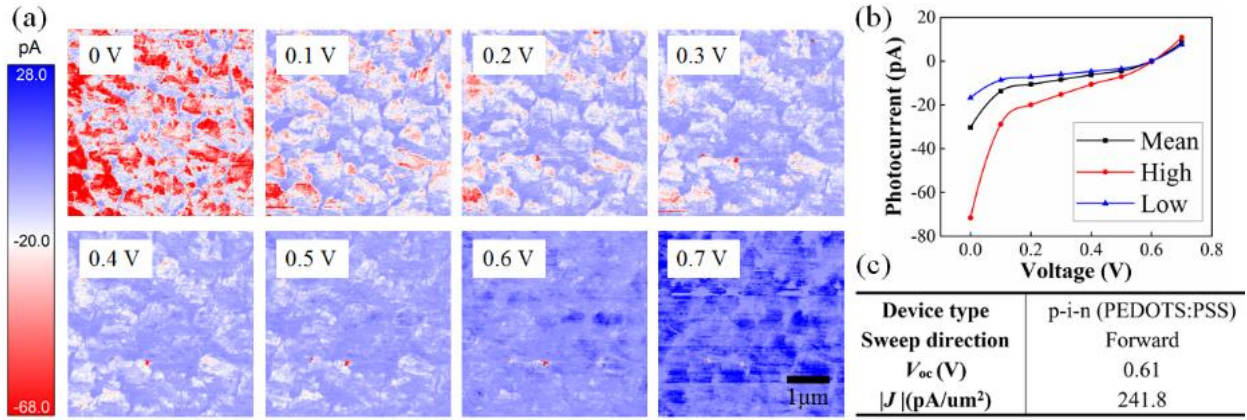

**Fig. 2** Mapping $CH_3NH_3PbI_3$ on FTO/PEDOTS:PSS by pcAFM; (a) photocurrent mapping under a series of DC biases; (b) approximated IV curves; (c) estimated open-circuit photovoltage and short-circuit photocurrent density. These data were unpublished, with sample prepared according to Ref. [17], and all pcAFM measurements were carried out on an Asylum Research MFP-3D AFM using ORCA conductive AFM with a gain of $5 \times 10^8$ V/A and a Nanosensors PPP-EFM conductive probe with a spring constant of 3.18 N/m. During the measurement, the sample was biased with specified DC voltage and illuminated in air through the bottom FTO cathode using an unfiltered CREE MK-R 12V LED.

## *Work function and dynamics of photocarriers*

Kelvin probe force microscopy (KPFM) is capable of detecting the contact potential difference between the sample surface and the conductive probe tip [18], and thus can be used to measure the work function of the sample if the tip's work function is known. In contrast to cAFM and pcAFM, KPFM is a local measurement that is only sensitive to the sample surface underneath the tip, and it has been widely used to study halide perovskite solar cells, such as variations of surface potential and work function across grain and grain boundaries, especially changes induced by the light illumination [19–21]. One example is shown in Fig. 3, revealing variation of surface potential distribution after light is turned on and off, wherein reversible evolution upon a sequence of dark scans is observed [22]. Such change is attributed to ionic migrations, for example negatively charged Pb and Methylammonium (MA) vacancies and positively charged I vacancy, which were also observed after electric poling [23]. Of particular interest is the dynamics of photocarries that occurs at much faster time scale, and it could be studied by more advanced techniques such as time-resolved EFM (TrEFM) [24], though it would be much more challenging to do. When KPFM is used to probe the cross-section of a halide perovskite solar cell, valuable information on the interfaces can also be mapped, including band bending and charge accumulation [25].

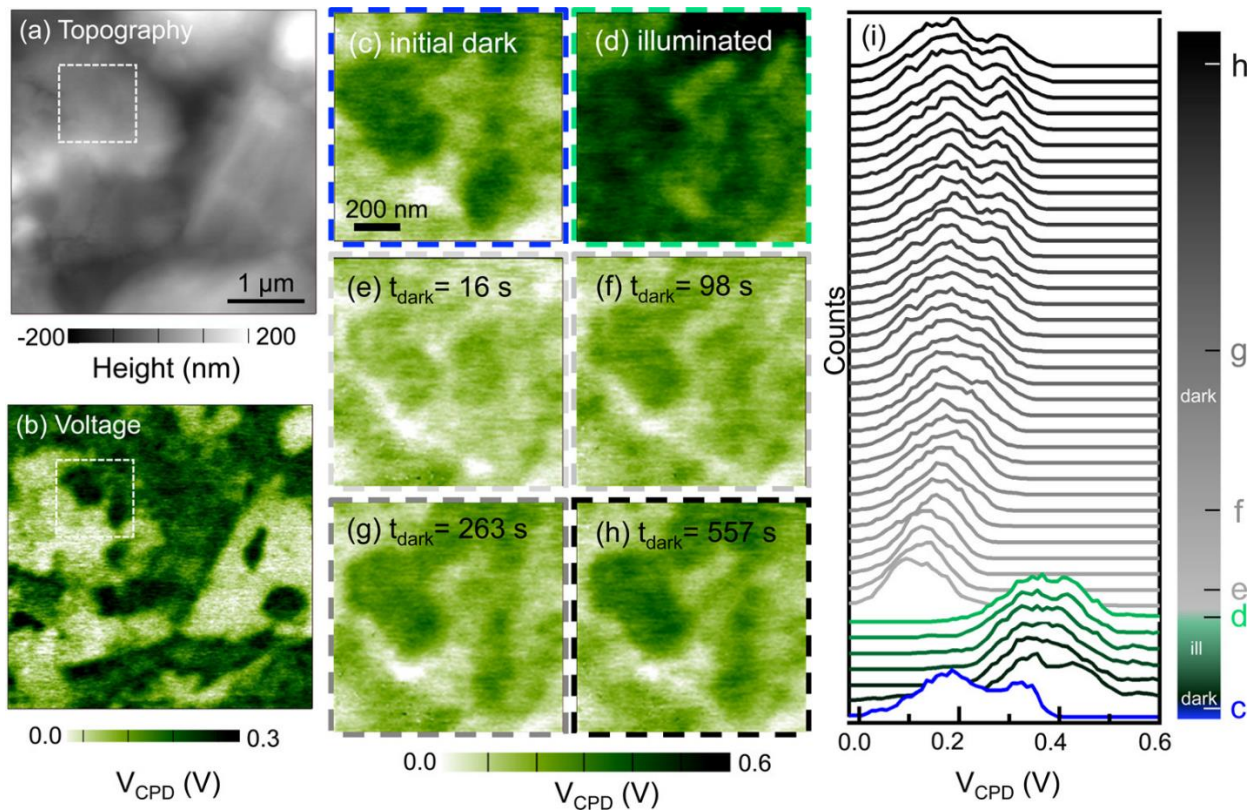

**Fig. 3** Surface potential evolution of perovskite solar cells at the nanoscale [22]; (a) topography and (b) dark-KPFM measurements on a representative area. A sequence of fast-KPFM measurements were acquired on the region highlighted by the white dashed square in (a,b): (c) dark-KPFM, (d) illuminated-KPFM, (e-h) dark-KPFM scans as a function of time, and (i) histograms of voltage distribution as a function of time showing reversible dynamics upon a sequence of dark scans. Reproduced with permission from American Chemical Society.

While KPFM has been used to examine ionic migration in halide perovskite as exhibited by surface potential variation, it is important to keep in mind that it relies on the long-range electrostatic interactions between the sample and the probe, and thus its spatial resolution is limited. This makes it difficult to probe truly local behavior, for example interplay between ion migration and defects, which can be imaged via Vegard strain with high sensitivity and spatial resolution [10,11], as discussed next. It is also noted that the principle of KPFM is based on aligning the vacuum levels of sample and tip (and thus separating their respective Fermi levels initially aligned at equilibrium), and the external potential can be applied to the probe or sample, either of which could be used as the reference point. Cautions thus must be excised in data analysis and interpretation to avoid confusions.

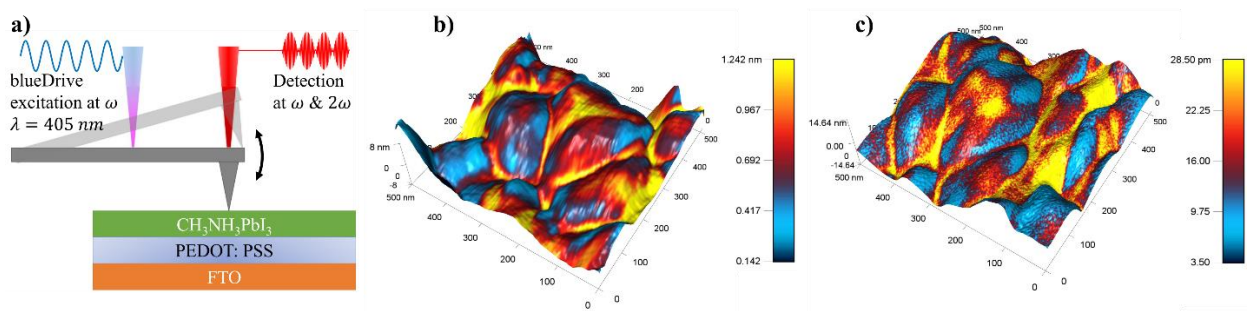

**Fig. 4** Ionic mapping of $CH_3NH_3PbI_3$ on FTO/PEDOTS:PSS by STIM; (a) schematics of STIM through photothermal heating, as detailed in Ref. [26]; (b,c) STIM mappings of (b) thermomechanical (first harmonic) and (c) ionic (second harmonic) responses, both overlaid on three-dimensional topography. These data were unpublished, with sample prepared according to Ref. [17], and STIM measurements were carried out on an Asylum Research Cypher ES AFM equipped with BlueDrive™ photothermal excitation unit that illuminates a 9mW power modulated 405nm laser aligned on the base of a gold coated probe (Multi75GD-G, Budget Sensors).

*Ions and ionic motions*

Ions and ionic motions involving charged Pb, MA ,and I vacancies are prominent in halide perovskites, especially under light illumination or electric loading, and it is thought to be the primary cause of IV hysteresis and contributes to the device degradation as well [27–30]. The phenomena, however, are often inferred from macroscopic measurement of transient current, and thus it is rather difficult to pinpoint the effects of microstructure. KPFM has been used to reveal surface potential redistribution associated with ionic migration [15], though the spatial resolution is limited. Imaging based on Vegard strain can in principle overcome this difficulty. The idea is to excite the concentration fluctuation of ionic species via scanning probe, resulting in corresponding oscillation of Vegard strain and displacement that can be mapped in contact mode at the nanoscale. One implementation is the electrochemical strain microscopy (ESM) that induces ionic fluctuation via charged scanning probe through AC voltage [31,32], and the other is the scanning thermo-ionic microscopy (STIM) that achieves the same via thermal stress [33], induced via either a heated probe or localized laser [26], as shown in Fig. 4(a). By examining the first and second harmonic responses of cantilever deflection induced by modulated laser intensity, both thermomechanical and Vegard strain responses can be obtained. Our analysis suggests that the ESM and STIM responses correlate with local ionic concentration and diffusivity [34–37], and thus can be used to map local ionic activities, as shown in Fig. 4(b,c) for $CH_3NH_3PbI_3$ on FTO/PEDOTS:PSS substrate [17], where enhanced responses at grain boundaries is evident, reflecting their higher ionic activities. Furthermore, relaxation behavior of ESM and STIM after DC perturbation is only governed by local diffusivity, making it possible to deconvolute diffusivity from the ionic concentration. How to distinguish the contributions of different ions and multiple types of vacancies, however, remains challenging, requiring careful analysis of their phase signal as well as relaxation behavior.

### *Polarization and ferroelectric switching*

Scanning thermo-ionic microscopy (STIM) is particularly suitable for detecting ionic activities in halide perovskites, since dynamic strain as measured by ESM may consist contributions from not only ionic fluctuation, but also piezoelectric strain, for which it is known as piezoresponse force microscopy (PFM) [11,38,39]. Indeed, halide perovskites are predicted to be ferroelectric [5], and polarization is suggested to be responsible for improved charge separation on the positive side, and contribute to IV hysteresis on the negative side. The unambiguous establishment of their ferroelectricity, however, is difficult, as leakage current due to ionic motions make reliable macroscopic measurement virtually impossible, which also smears the PFM signal at the microscopic scale [40]. Furthermore, different processing routes and conditions from different groups often result in different microstructures and even phases, making direct and reliable comparison of respective ferroelectric characterizations difficult. Indeed, there have been numerous reports claiming that $CH_3NH_3PbI_3$ is not ferroelectric [41]. Nevertheless, it has recently been demonstrated that the electromechanical strain in at least some of $CH_3NH_3PbI_3$ is dominated by the first harmonic response, and thus could be linear piezoelectric [17], and clear domain structures were also revealed in a number of studies [41–44], as shown in Fig. 5(a), with some of them exhibiting characteristic switching behavior, as shown in Fig. Fig. 5(b). Substantial shifting in nucleation bias induced by lightening is also revealed, as summarized in Fig. 5(c), resulting in strong switching asymmetry and sometimes photo-induced switching [17]. These data suggest that $CH_3NH_3PbI_3$ with appropriate microstructure and phase could be ferroelectric, though macroscopic as well as atomic scale evidence are still needed in order to establish the phenomenon without ambiguity. The implication of possible ferroelectric polarization on photovoltaic performance of halide perovskites also requires further studies.

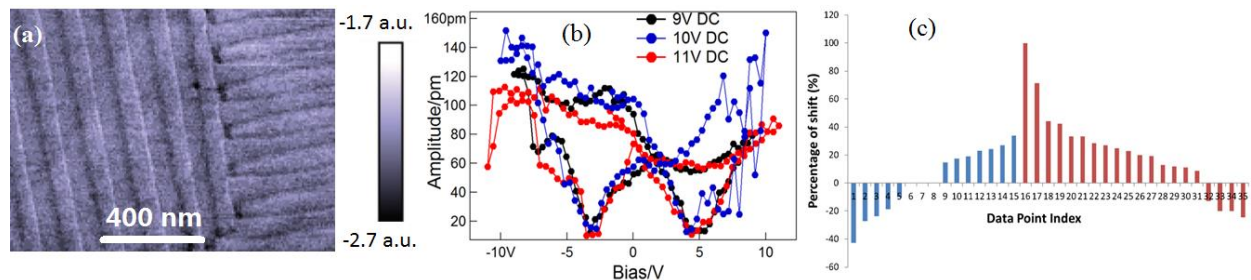

**Fig. 5** Piezoresponse force microscopy (PFM) of $CH_3NH_3PbI_3$; (a) PFM mapping showing characteristic lamellar domain structure [44]; (b) PFM butterfly loop and (c) coercive voltage asymmetry, defined as $(V_+ + V_-)/(V_+ - V_-)$ wherein $V_+$ and $V_-$ are positive and negative coercive voltage, before (blue) and after (red) light illumination [17], showing that substantial switching asymmetry is induced by lightening. Reproduced with permission from Royal Society of Chemistry.

*Composition heterogeneity and degradation*

Besides functional properties, SPM can also be used to interrogate the composition heterogeneity of halide perovskites when combined with either Raman or infrared (IR) spectroscopy, and it is especially valuable for studying their degradations. As schematically shown in Fig. 6(a), a pulsed IR laser induces abrupt and short-lived thermal expansion in the sample under atomic force microscopy-based infrared spectroscopy (AFM-IR) [45], initially proposed by Alex Dazzi, which is also similar to STIM we discussed earlier. The rapid thermal expansion in turn excites $2^{nd}$ harmonic resonant oscillation of the cantilever, whose amplitude is directly proportional to the sample absorption coefficient, making it possible to probe chemical heterogeneity and local IR spectrum with 10 nm resolution, 3 orders of magnitude higher than standard FTIR. The technique has recently been used to study halide perovskite solar cells, including chloride incorporation process [46] and ferroelastic domains [41], and we have carried out AFM-IR to study the degradation of $CH_3NH_3PbI_3$ on FTO/PEDOTS:PSS substrate as well [17]. As seen in Fig. 6(b), the fresh sample shows a dominant peak at 1464 cm$^{-1}$ in AFM-IR spectrum, corresponding to $CH_3$ antisymmetric stretching of the MA ion. The sample degraded by air exposure, however, possesses additional peaks at 1296 and 1264 cm$^{-1}$ (Fig. 6(c)), corresponding to OH bond that results from reaction between $CH_3NH_3PbI_3$ and water in air [8,9,47,48]. Furthermore, the chemical mappings for these two samples under 1464 cm$^{-1}$ adsorption reveal distinct features, as evident in Fig. 6(de), wherein degraded sample shows much weaker adsorption with more prominent heterogeneity, again indicating reactions with water.

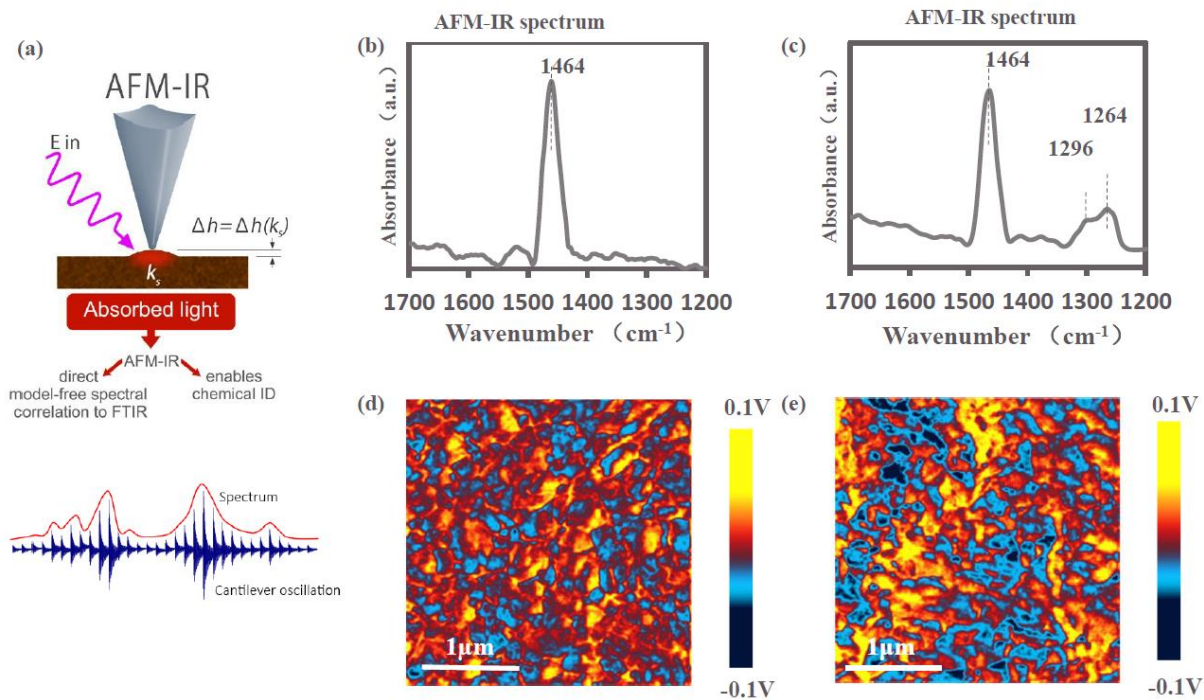

**Fig. 6** Chemical heterogeneity of $CH_3NH_3PbI_3$ on FTO/PEDOTS:PSS revealed by AFM-IR; (a) schematics of working principle, as detailed in Ref. [45]; (b) and (c) spectra of fresh and degraded sample; (d) and (e) chemical mapping of fresh and degraded sample at 1464cm$^{-1}$, respectively. These data were unpublished. AFM-IR measurements were carried out on an Anasys Instruments nano-IR2 AFM equipped with one Daylight solutions pulsed QCL laser ranged from 1200-1700cm$^{-1}$, using Anasys Instruments PR-EX-TnIR cantilever having spring constant of 2nN/nm). AFM topography was scanned firstly using tapping mode and then a serial of locations were selected for contact-mode AFM-IR spectrum measurement. The laser power was set as 3% and 270kHz pulse rate were optimized to maximize the IR signal. After the local IR spectrum is collected, the wavelength is fixed at 1464cm$^{-1}$, and then tapping AFM-IR image was collected at this wavelength to get the distribution of 1464cm$^{-1}$ band. The fresh sample was prepared in advance according to Ref. [17] and sealed in cell filled with $N_2$ protective atmosphere, and then taken out for immediate AFM-IR measurement; the degraded sample was exposed in air for 5 hours before AFM-IR measurement.

**Outlook**

It is evident that functional probing and imaging of perovskite solar cells have shed considerable insight into local photovoltaic characteristics of the organometal halide perovskites and their correlation with the microstructure, yet there is a long way to go to fully realize the promises and potentials offered by SPM. Here, we discuss some future directions for further studies.

First of all, multi-modality imaging is essential for a comprehensive understanding of the local behavior in halide perovskites. Each of the technique discussed above only focuses on a particular aspect of the local heterogeneity and inhomogeneous response, such as photocurrent, surface potential, mechanical displacement, or chemical heterogeneity, yet multiple microstructural mechanisms may contribute to such responses. For example, surface potential can be shifted by photocarriers, ionic motions, charge trapping, or polarization switching, while mechanical displacement can be induced by piezoelectricity, ionic fluctuation, or electrostatic interactions. Thus it would be impossible to trace down the dominant mechanisms using any of the techniques alone, and multiple techniques must work in concert on the same scanned area, especially for connecting chemical heterogeneity probed by AFM-IR with local responses by other SPM techniques.

Secondly, in-operando SPM is critical for revealing microscopic mechanisms that are responsible for the macroscopic performance of halide perovskite solar cells. This requires mapping the local response under global perturbation, for example during measurements of IV curves or impedance, which would make it possible to monitor critical heterogeneities such as interfaces, grain boundaries, and defect sites during device operation. While some work along this direction has been developed, for example under light illumination or electric poling, much more remains to be explored, especially mapping under a series of global perturbation with different frequencies or voltages. This would require further development of hardware instrumentation as well as analytic software.

Thirdly, multi-dimensional data will be generated by SPM scans, especially with multi-modality imaging and in-operando scanning, and advanced data analysis is invaluable. For example, in order to achieve high resolution IV characteristics under pcAFM, sequence of images have to be acquired under different DC biases, and potentially under different lighting conditions as well. Multivariate statistical tools such as principal component analysis (PCA) is powerful in extracting clear physical insight from noisy data [49,50], as shown by the example in Fig. 7. Through orthogonal transformation, PCA converts the set of possibly correlated variables, in this case pcAFM mappings of Fig. 2 under different DC biases, to a set of linearly uncorrelated variable known as principal components, represented by the corresponding loading mappings in Fig. 7(c,d). The principal components are sorted by their eigenvalues in a descending manner, with the first principal component accounting for the maximum possible variability in the data, as shown by the scree plot in Fig. 7(a). The corresponding first two eigenvectors are shown in Fig. 7(b), revealing the trend of the first two orthogonal components as a function of DC bias. Physically, the pcAFM mappings under different biases contain two important information, the variation of the photocurrent density with respect to the spatial locations and with respect to applied biases, which are interconnected in the original mappings of Fig. 2. Under PCA, the data are reorganized, such that the spatial variation is best represented by the loading mappings, and the bias dependence is reflected in the eigenvectors. Since the dimensionality of the data is reduced by PCA and hence the components with low variance such as noises are removed, the single-to-noise ratios are enhanced, as evident in Fig. 7(c). Due to the unsupervised nature of PCA, it is often not straightforward to interpret the physical meaning of each PCA mode, as they are obtained solely through a statistical procedure. Nevertheless, we believe the first mode mapping in Fig. 7(c) accounts for spatial variation of the photocurrent density, and the second mode mapping in Fig. 7(d), which has an order of magnitude lower variance, represents the fluctuation of the data. Meanwhile, the first eigenvector variation shown by Fig. 7(b) can be viewed as representing *IV* characteristics of the photocurrent. Further details on PCA can be found in our recent publication [51].

Finally, we call for more close collaboration on perovskite solar cells among material scientists, device engineers, scanning probe microscopists, and data scientists. We believe such multidisciplinary interactions are invaluable for fully understanding these complex systems, and will help to speed up the development of viable commercial devices. As evident from glimpses of recent SPM studies on halide perovskite solar cells highlighted in this perspective, much remain to be learned on their microscopic processes, including truly dynamic behavior of photocarriers, transport characteristics of different types of charged ions and vacancies, and significance of ferroelectric polarizations. Scanning probe microscopies alone are definitely limited in their capabilities, yet they can be very powerful when working together with other techniques in concert.

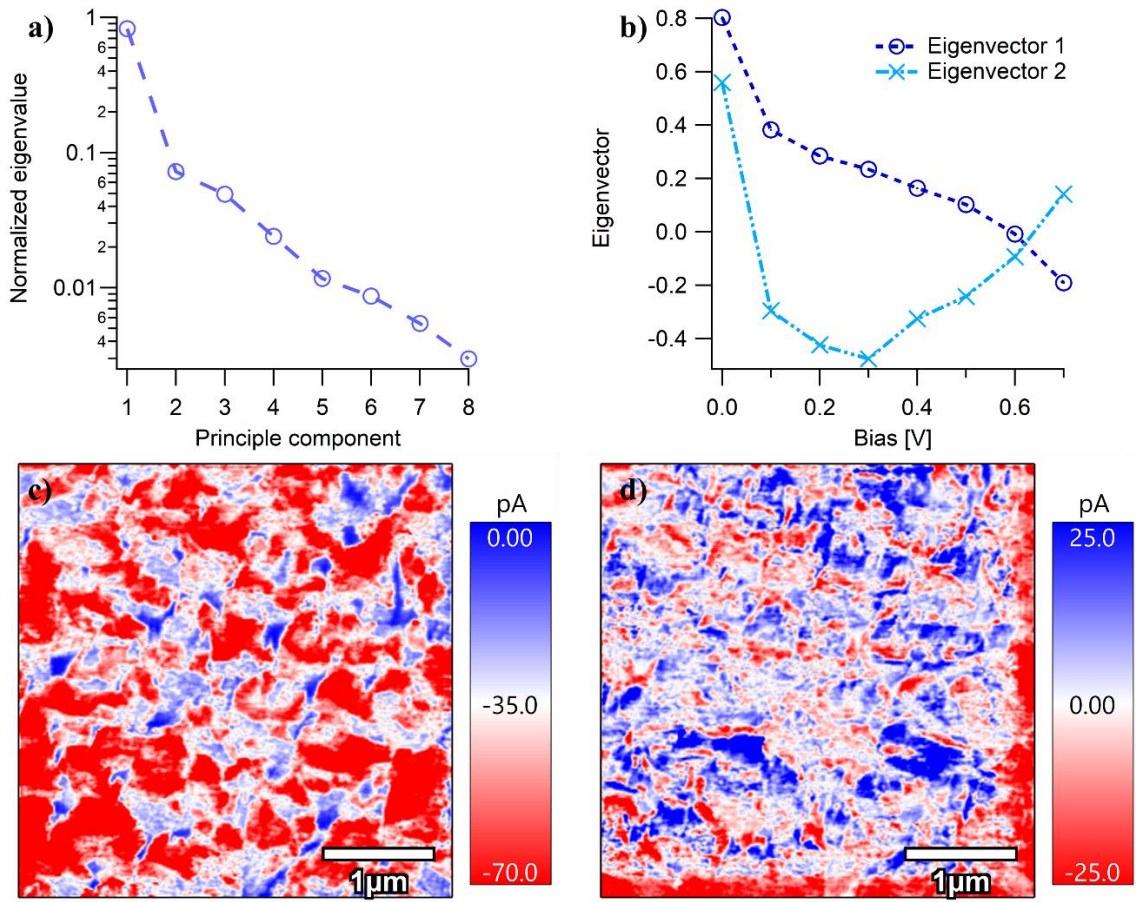

**Fig. 7** Multivariate statistical analysis of pcAFM mappings of Fig. 2 via PCA; (a) PCA scree plot of normalized eigenvalues versus principal components (PC) showing that the first mode accounts for the maximum possible variability in the data; (b) eigenvectors for PC mode 1 and 2 showing IV characteristics; (c) and (d) associated spatial mapping of mode 1 and 2 showing improved signal-noise ratio compared to Fig. 2. These data were unpublished, and the data analysis method and post-processing scripts are available in section 2.2 of our recent publication [51].

**Acknowledgements**

We acknowledge National Key Research and Development Program of China (2016YFA0201001), US National Science Foundation (CBET-1435968), National Natural Science Foundation of China (11627801, 11472236, and 51102172), the Leading Talents Program of Guangdong Province (2016LJ06C372), Shenzhen Knowledge Innovation Program (JCYJ20170307165905513), and Key Laboratory for Magnetic Resonance and Multimodality Imaging of Guangdong Province (2014B030301013). This material is based in part upon work supported by the State of Washington through the University of Washington Clean Energy Institute. LW and JY also thank the Lab of Interfacial Electrochemistry for Energy Materials led by Prof. Wenbin Cai in the Department of Chemistry, Fudan University, which provided access to nanoIR2-fs system for AFM-IR measurement.

**Competing Interests**

There is no competing interests.

**Contributions**

JL initiated the project with instrumental help from JZ and WC, and wrote the manuscript; BH organized majority of the literature, and produced data for Fig. 2; EE produced data for Fig. 4, and carried out PCA for Fig. 7; LW and JY produced data for Fig. 6; JZ provided all the samples for the testing; and all the authors read the manuscript and participated the revision.